\documentclass[aps,prd,preprint,showpacs,groupaddress,superscriptaddress,nofootinbib]{revtex4}
\usepackage{color}
\usepackage{graphicx}
\usepackage{tabularx}
\usepackage{amsmath}
\usepackage{amssymb}
\usepackage{hyperref}

\begin{document}

\title{Gauge-singlet dark matter in a left-right symmetric model 
with spontaneous $CP$ violation}

\author{ Wan-Lei Guo\footnote{guowl@itp.ac.cn}}
\affiliation{Kavli Institute for Theoretical Physics China,
Institute of Theoretical Physics, \\ Chinese Academy of Science,
Beijing 100080, P.R.China}

\author{ Li-Ming Wang\footnote{wanglm@itp.ac.cn}}
\affiliation{Kavli Institute for Theoretical Physics China,
Institute of Theoretical Physics, \\ Chinese Academy of Science,
Beijing 100080, P.R.China}

\author{ Yue-Liang Wu\footnote{ylwu@itp.ac.cn}}
\affiliation{Kavli Institute for Theoretical Physics China,
Institute of Theoretical Physics, \\ Chinese Academy of Science,
Beijing 100080, P.R.China}

\author{Yu-Feng Zhou\footnote{yfzhou@kias.re.kr}}
\affiliation{Korea Institute for Advanced Study, Seoul 130-722,
Korea}

\affiliation{Kavli Institute for Theoretical Physics China,
Institute of Theoretical Physics, \\ Chinese Academy of Science,
Beijing 100080, P.R.China}

\author{Ci Zhuang\footnote{zhuangc@itp.ac.cn}}
\affiliation{Kavli Institute for Theoretical Physics China,
Institute of Theoretical Physics, \\ Chinese Academy of Science,
Beijing 100080, P.R.China}

\begin{abstract}
We propose a dark matter (DM) scenario in an extension of a left-right
symmetric model with a gauge-singlet scalar field. The gauge-singlet scalar
can automatically become a DM candidate, provided that both $P$ and
$CP$ symmetries are only broken spontaneously.  Thus no extra discrete
symmetries are needed to make the DM candidate stable.
After constraining the model parameters from the observed relic DM
density we make predictions for direct detection experiments. We show
that for some parameter range, the predicted WIMP-nucleon elastic
scattering cross section can reach the current experimental upper
bound, which can be tested by the experiments in the near future.

\end{abstract}

\pacs{95.35.+d, 12.60.-i}

\preprint{KIAS-P08078}

\maketitle

\section{Introduction}

The Standard Model (SM) of particle physics, although greatly
successful in phenomenology gives no explanations for parity ($P$) and
$CP$ violation.  The observed neutrino oscillations, the large
baryonumber asymmetry and large energy density from non-baryonic dark
matter (DM) in the universe are clear indications for new physics
beyond the SM.
In the left-right (LR) symmetric models for weak
interactions \cite{Pati:1974yy,Mohapatra:1974hk,Senjanovic:1975rk,Deshpande:1990ip}, the
left- and right-handed fermions are treated equally, and $P$ symmetry
is restored prior to the spontaneous symmetry breaking (SSB). The LR
models have other advantages such as the gauge group $SU(2)_L\otimes
SU(2)_R\otimes U(1)_{B-L}$ can be elegantly embedded into grand
unification theories, and the right-handed neutrinos are naturally
required, etc..

The LR models may also contain DM candidates. In one of the minimal
versions of the LR model which contains one Higgs bidoublet $\phi$ and
two Higgs triplets $\Delta_{L,R}$, a $Z_2$ symmetry on the left triplet
$\Delta_L\to -\Delta_L$ can be used to resolve  the so-called VEV
seesaw problem \cite{Deshpande:1990ip}. A direct
consequence of this discrete symmetry is that the neutral component
$\delta^0_L$ of $\Delta_L $ can only annihilate or be produced by
pairs, which makes $\delta^0_L$ a potential DM candidate. However, due
to the fact that $\delta^0_L$ participates $SU(2)_L$ gauge
interactions which is quite strong, the WIMP-nucleon elastic
scattering experiments lead to a severe constraint on the dark matter
relic density which is a few order of magnitudes below the observed
value \cite{Guo:2008hy}.  Thus it cannot be a main source of DM in the
universe.

A disadvantage of the minimal LR model is that although $P$ can be broken
spontaneously, the $CP$ symmetry has to be broken explicitly, which looks
quite unnatural. The reason is that without large fine-tuning on the Higgs
potential the condition for spontaneous $CP$ violation (SCPV) cannot be satisfied
\cite{Basecq:1985sx,Masiero:1981zd,Deshpande:1990ip}. Furthermore, in the
minimal LR model with SCPV the predicted $CP$ phase angle
$\sin2\beta\sim 0.1$ in the unitarity triangle of the
Cabibbo-Kobayashi-Maskawa (CKM) matrix is far bellow the experimentally
measured value of $\sin2\beta=0.671\pm 0.024$ from the two B-factories
\cite{Ball:1999mb}. The minimal LR model also suffers from strong
phenomenological constraints from low energy flavor-changing-neutral-current
(FCNC) processes, especially the neutral kaon mixing which push the masses of
the right-handed gauge bosons and some neutral Higgs bosons above the TeV
scale
\cite{Mohapatra:1983ae,Gilman:1983ce,Chang:1984hr,Hou:1985ur,Basecq:1985cr,Ecker:1985vv,Basecq:1988cv}.

Motivated by the requirement of both spontaneous $P$ and $CP$
violation, We have discussed an extension of the minimal LR model with
two Higgs bidoublets \cite{Wu:2007kt,Wu:2007gb,wang2009}. In this two-Higgs-bidoublet 
LR model (2HBLR), the additional Higgs bidoublet may modify the 
Higgs potential such that the fine-tuning problem can be avoided.  
The extra Higgs bidoublet can also change the interferences among different
contributions in the box-diagrams in the
neutral meson mixings, and lower the bounds for right-handed gauge boson masses
to be bellow the TeV scale \cite{Wu:2007kt,Wu:2007gb}.

Note that the spontaneous $P$ and $CP$ violation in the LR models can
also be useful for DM model-building. Before the SSB, the Lagrangian for
the particle interactions prohibits the $P$-odd and $CP$-odd
interactions, which may prevent the decays of the particles with odd
$CP$ parity. These particles can remain stable even after the SSB,
provided that they do not develop VEVs and do not couple to the
symmetry breaking sector. The simplest case would be that there is a
gauge-singlet scalar field with odd $CP$ parity, and has a vanishing
VEV.

In this work we discuss this possibility by considering an
extension of the 2HBLR with a gauge-singlet complex field $S$ which plays
the role of DM candidate, and  the stability of DM is purely protected 
by the discrete $P$ and $CP$ symmetries.
This model distinguishes itself from the previous gauge-singlet models ( see, e.g. 
\cite{McDonald:1993ex,Bento:2000ah,Burgess:2000yq,Davoudiasl:2004be,Kusenko:2006rh,
Barger:2007im,Barger:2008jx,He:2008qm} )
in that no {\it ad hoc}  discrete symmetry of $Z_2$ type is introduced.
This possibility has not been emphasized before in the literature,
simply because most of the popular models such as SM and MSSM violate
$P$ and $C$ maximally. This simple model shows that the DM may be
connected to the fundamental symmetries of the quantum field
theory. Recently, it is also noticed that the custodial symmetry of
the gauge interaction can also be used to stabilize the DM
candidate \cite{Hambye:2008bq}.
We calculate in this model the cross sections for the DM annihilation and
the elastic scattering with the nucleus. The results show that for a
large parameter space the DM relic density can be reproduced.  The
correlation between the DM annihilation  and the elastic scattering off
the nucleus depends on the Higgs and Yukawa sector of the model, and can
be quite different from the ordinary gauge-singlet model which in some 
limit has a simple one to one correspondence. In general, the predictions
for the direct detection experiments can be significantly larger and 
can even reach the current experimental upper bound for large Yukawa couplings.

This paper is organized as follows: In Section. \ref{model}, we
outline the main feature of the model. In Section \ref{DM-in-model},
we discuss the parameter space, and give the formulas for main
processes for the DM annihilation and the elastic scattering off the
nucleus in a simplified case where one Higgs bidoublet decouples from
the theory. The case in which both Higgs bidoublet are active is discussed 
in Section \ref{DM-2HBLR}. We finally conclude in Section \ref{conclusions}.

\section{The LR symmetric model with a gauge-singlet}\label{model}

We begin with a LR model with a gauge group 
$SU(2)_L\otimes SU(2)_R\otimes U(1)_{B-L}$
which contains  two Higgs bidoublets
$\phi$ (2,$2^{*}$,0), $\chi$ (2,$2^{*}$,0), a left(right)-handed Higgs triplet $\Delta_{L(R)}$ (3(1),1(3),2),
and a gauge-singlet $S$(0,0,0) with the following flavor contents
\begin{eqnarray}
\phi  = \left ( \begin{matrix} \phi_1^0 & \phi_1^+ \cr \phi_2^- &
\phi_2^0 \cr  \end{matrix} \right ) , \; 
\chi  = \left ( \begin{matrix} \chi_1^0 & \chi_1^+ \cr \chi_2^- &
\chi_2^0 \cr  \end{matrix} \right ) , \; 
\Delta_{L,R}  = \left (
\begin{matrix} \delta_{L,R}^+/\sqrt{2} & \delta_{L,R}^{++} \cr
\delta_{L,R}^{0} & -\delta_{L,R}^{+}/\sqrt{2} \cr \end{matrix}
\right ) , \;
S=\frac{1}{\sqrt2}(S_\sigma+i S_D) . 
\nonumber\\
\end{eqnarray}
The introduction of Higgs bidoublet $\chi$ is to overcome the problem of
fine-tuning in generating the SCPV in the minimal LR model,
and relax the severe low energy phenomenological constraints \cite{Wu:2007kt,Wu:2007gb,wang2009}.
Under the $P$ and $CP$ transformation, these fields transform as
\begin{eqnarray}
\begin{tabular}{lll}\hline
  &  $P$  &   $CP$\\ \hline
$\phi$ \;   &  $\phi^\dagger$ \; &$\phi^*$  
\\ 
$\chi$ \;   &  $\chi^\dagger$ \; &$\chi^*$ 
\\
$\Delta_{L(R)}$ \;  & $\Delta_{R(L)}$ \; & $\Delta_{L(R)}^*$
\\
$S$  &  $S$ &   $S^*$
\\ \hline
\end{tabular}
\label{pcp}
\end{eqnarray}
We shall require $P$ and $CP$ invariance of the Lagrangian, which strongly restricts
the structure  of the Higgs potential. For instance, for the terms 
involving the $\phi$ and $\Delta_{L,R}$ fields the most general
potential is given by
\begin{eqnarray}
-\mathcal{V}_{\phi\Delta} & = & -\mu_{1}^{2}\rm{Tr}(\phi^{\dagger}\phi)-\mu_{2}^{2}[\rm{Tr}(\tilde{\phi}^{\dagger}\phi)+\rm{Tr}(\tilde{\phi}\phi^{\dagger})]-\mu_{3}^{2}[\rm{Tr}(\Delta_{L}\Delta_{L}^{\dagger})+\rm{Tr}(\Delta_{R}\Delta_{R}^{\dagger})]\nonumber\\
 &  & +\lambda_{1}[\rm{Tr}(\phi^{\dagger}\phi)]^{2}+\lambda_{2}^{2}\{[\rm{Tr}(\tilde{\phi}^{\dagger}\phi)]^{2}+[\rm{Tr}(\tilde{\phi}\phi^{\dagger})]^{2}\}+\lambda_{3}[\rm{Tr}(\tilde{\phi}^{\dagger}\phi)\rm{Tr}(\tilde{\phi}\phi^{\dagger})]\nonumber\\
 &  & +\lambda_{4}\{\rm{Tr}(\phi^{\dagger}\phi)[\rm{Tr}(\tilde{\phi}^{\dagger}\phi)+\rm{Tr}(\tilde{\phi}\phi^{\dagger})]\}\nonumber\\
 &  & +\rho_{1}\{[\rm{Tr}(\Delta_{L}\Delta_{L}^{\dagger})]^{2}+[\rm{Tr}(\Delta_{R}\Delta_{R}^{\dagger})]^{2}\}+\rho_{2}[\rm{Tr}(\Delta_{L}\Delta_{L})\rm{Tr}(\Delta_{L}^{\dagger}\Delta_{L}^{\dagger})+\rm{Tr}(\Delta_{R}\Delta_{R})\rm{Tr}(\Delta_{R}^{\dagger}\Delta_{R}^{\dagger})]\nonumber\\
 &  & +\rho_{3}[\rm{Tr}(\Delta_{L}\Delta_{L}^{\dagger})\rm{Tr}(\Delta_{R}\Delta_{R}^{\dagger})]+\rho_{4}[\rm{Tr}(\Delta_{L}\Delta_{L})\rm{Tr}(\Delta_{R}^{\dagger}\Delta_{R}^{\dagger})+\rm{Tr}(\Delta_{L}^{\dagger}\Delta_{L}^{\dagger})\rm{Tr}(\Delta_{R}\Delta_{R})]\nonumber\\
 &  & +\alpha_{1}\rm{Tr}(\phi^{\dagger}\phi)\rm{Tr}(\Delta_{L}\Delta_{L}^{\dagger})+\rm{Tr}(\Delta_{R}\Delta_{R}^{\dagger})] 
          +\alpha_{2}Tr[ (\tilde{\phi}^{\dagger}\phi)+(\tilde{\phi}\phi^{\dagger})]Tr[(\Delta_{L}\Delta_{L}^{\dagger})+(\Delta_{R}\Delta_{R}^{\dagger})]\nonumber\\
 &  & +\alpha_{3}[\rm{Tr}(\phi\phi^{\dagger}\Delta_{L}\Delta_{L}^{\dagger})+\rm{Tr}(\phi^{\dagger}\phi\Delta_{R}\Delta_{R}^{\dagger})]\nonumber\\
 &  & +\beta_{1}[\rm{Tr}(\phi\Delta_{R}\phi^{\dagger}\Delta_{L}^{\dagger})+\rm{Tr}(\phi^{\dagger}\Delta_{L}\phi\Delta_{R}^{\dagger})]+\beta_{2}[\rm{Tr}(\tilde{\phi}\Delta_{R}\phi^{\dagger}\Delta_{L}^{\dagger})+\rm{Tr}(\tilde{\phi}^{\dagger}\Delta_{L}\phi\Delta_{R}^{\dagger})]\nonumber\\
 &  & +\beta_{3}[\rm{Tr}(\phi\Delta_{R}\tilde{\phi}^{\dagger}\Delta_{L}^{\dagger})+\rm{Tr}(\phi^{\dagger}\Delta_{L}\tilde{\phi}\Delta_{R}^{\dagger})] ,
\label{Vphidelta}
\end{eqnarray}
where the coefficients $\mu_i$, $\lambda_i$, $\rho_i$, $\alpha_i$ and
$\beta_i$ in the potential are all real as all the terms are
self-Hermitian. The Higgs potential $\mathcal{V}_{\chi\Delta}$
involving $\chi$ field can be obtained by the replacement
$\chi\leftrightarrow \phi$ in Eq. (\ref{Vphidelta}). The mixing term
$\mathcal{V}_{\chi\phi\Delta}$ can be obtained by replacing one of
$\phi$ by $\chi$ in all the possible ways in Eq. (\ref{Vphidelta}).

In order to simplify the discussion,  in this section we shall first
consider a simple case in which the bidoublet $\chi$ does not
mix significantly with other fields. In this case the model is reduced
to the minimal LR model plus a gauge-singlet, which already
contains the main features of the complete model. We postpone the
discussions on the $\chi$ contributions into Section \ref{DM-2HBLR}. The
most general Higgs potential involving the singlet field $S$ is given
by
\begin{eqnarray}
-\mathcal{V}_{S}&=&\frac{1}{\sqrt{2}}\tilde\mu_0^3(S+S^*)-\tilde\mu_S^2SS^*-\frac{1}{4}\tilde\mu_{\sigma}^2(S+S^*)^2
+\sqrt{2}\tilde\mu_{\sigma S}(S+S^*)SS^*\nonumber\nonumber\\
&&+\frac{1}{6\sqrt{2}}\tilde\mu_{3\sigma}(S+S^*)^3
+\tilde\lambda_S(SS^*)^2-\frac{1}{4}\tilde\lambda_{\sigma S}(S+S^*)^2SS^*-\frac{1}{16}\tilde\lambda_{\sigma}(S+S^*)^4\nonumber\\
&&+\sum_{i=1}^3\left[-\frac{1}{\sqrt{2}}\tilde\mu_{i,\sigma}(S+S^*)+\tilde\lambda_{i,S}SS^*
-\frac{1}{4}\tilde\lambda_{i,\sigma}(S+S^*)^2\right]O_i \, ,
\label{VS}
\end{eqnarray}
where
\begin{eqnarray}
O_1={\rm{Tr}}(\phi^{\dag}\phi),
O_2={\rm{Tr}}(\phi^{\dag}\tilde\phi+\tilde\phi^{\dag}\phi)
 \mbox{ and }
O_3={\rm{Tr}}(\Delta_L^{\dag}\Delta_L+\Delta_R^{\dag}\Delta_R)
\,.
\label{operators}
\end{eqnarray}
Note that it only involves combinations of $(S+S^*)$ and $SS^*$. The terms
proportional to odd powers of $(S-S^*)$ are absent in the singlet
self-interactions as they are $P$-even but $C$-odd. Furthermore, they cannot
mix with the Higgs multiplets in $O_i$ because the three independent
gauge-invariant combinations $O_i (i=1,\dots,3)$ in Eq. (\ref{operators}) are
both $P$- and $C$-even. Other possible Higgs multiplet combinations such as
${\rm{Tr}}(\phi^{\dag}\tilde\phi-\tilde\phi^{\dag}\phi)$ and
${\rm{Tr}}(\Delta_L^{\dag}\Delta_L-\Delta_R^{\dag}\Delta_R)$ are $P$-odd, thus
cannot couple to $S$.  The terms proportional to even powers of $(S-S^*)$ can
be rewritten in terms of $(S+S^*)^2$ and $SS^*$.  We have checked that the $P$
and $CP$ transformation rules for $S$ defined Eq.  (\ref{pcp}) is acturally
the only possible way for the implementation of the DM candidate.  For future
convenience we rewrite $\mathcal{V}_S$ in terms of the component field
$S_\sigma$ and $S_D$.
\begin{eqnarray}
-\mathcal{V}_{S}&\equiv&\mu_0^3S_{\sigma}-\frac{1}{2}\mu_{\sigma}^2S_{\sigma}^2-\frac{1}{2}\mu_{D}^2S_D^2
+\frac{1}{3}\mu_{3\sigma}S_{\sigma}^3+\mu_{\sigma
D}S_{\sigma}\,S_D^2
+\frac{1}{4}\lambda_{\sigma}S_{\sigma}^4+\frac{1}{4}\lambda_{D}S_D^4+\frac{1}{2}\lambda_{\sigma D}S_{\sigma}^2S_D^2\nonumber\\
&&+\sum_{i=1}^3(-\mu_{i,\sigma}S_{\sigma}+\frac{\lambda_{i,\sigma}}{2}S_{\sigma}^2+\frac{\lambda_{i,D}}{2}S_D^2)O_{i}\,,\label{potential}
\label{VsComponents}
\end{eqnarray}
with the redefined coefficients
\begin{align}
\mu_0 &=\tilde\mu_0,   &\mu_{\sigma}^2=&\tilde\mu_S^2+\tilde\mu_{\sigma}^2, &\mu_D^2 =&\tilde\mu_S^2,
\nonumber\\
\mu_{3\sigma} &=\tilde\mu_{3\sigma}+3\tilde\mu_{\sigma  S}, &\mu_{\sigma D} &=\tilde\mu_{\sigma S},
\nonumber\\
\lambda_{\sigma}&=\tilde\lambda_S-\tilde\lambda_{\sigma  S}-\tilde\lambda_{\sigma}, &\lambda_D&=\tilde\lambda_S, &\lambda_{\sigma  D}&=\tilde\lambda_{S}-\frac{1}{2}\tilde\lambda_{\sigma S},
\nonumber\\
\mu_{i,\sigma} &=\tilde\mu_{i,\sigma},&\lambda_{i,\sigma}&=\tilde\lambda_{i,S}-\tilde\lambda_{i,\sigma}, &\lambda_{i,D}&=\tilde\lambda_{i,S}.
\end{align}
It follows from Eq. (\ref{VsComponents}) that $S_D$ can only be
produced by pairs, thus  is a potential dark matter candidate.
After the SSB, the Higgs multiplets
obtain  nonzero VEVs
\begin{equation}
\langle\phi_{1,2}^0\rangle=\frac{\kappa_{1,2}}{\sqrt2}
\; \mbox{ and }\;
\langle\delta_{L,R}^0\rangle=\frac{v_{L,R}}{\sqrt2}
\end{equation}
where $\kappa_1$, $\kappa_2$, $v_L$ and $v_R$ are in general complex,
and $ \kappa \equiv \sqrt{|\kappa_1|^2 + |\kappa_2|^2} \approx 246$ GeV represents the
electroweak symmetry breaking (EWSB) scale. The value of $v_R$
sets the scale of LR symmetry breaking which is directly linked to the 
right-handed gauge boson masses.

With the extra contributions from $\mathcal{V}_S$ to the whole Higgs
potential $\mathcal{V}\equiv\mathcal{V}_{\phi\Delta}+\mathcal{V}_S$,
one needs to redo the minimization with respect to $\phi$ and
$\Delta_{L,R}$. However, from Eq. (\ref{Vphidelta}) and (\ref{VS}) it
follows that the minimization conditions for $\phi$ and
$\Delta_{L(R)}$ remain to have the same form as that in the minimal LR
model. This is because the mixing introduced by the singlet $S$ only
changes the overall coefficients $\mu_1$, $\mu_2$ and $\mu_3$ of the
$\phi$ and $\Delta_{L,R}$ potential term in Eq. (\ref{Vphidelta}).
Hence the mass matrix of the Higgs multiplet $\phi$ and $\Delta_{L,R}$
remains  the same as that in the minimal LR model in
Refs. \cite{Deshpande:1990ip,Duka:1999uc}, which also indicates that
the additional potential term $\mathcal{V}_S$ in Eq. (\ref{potential})
does not help in resolving the fine-tuning problem. The fine-tuning
can only be relaxed by introducing another bidoublet $\chi$.
From  the minimization condition  for the singlets $S_{\sigma}$, one can
eliminate one parameter $\mu_\sigma$  
\begin{eqnarray}
\mu_{\sigma}^2&=&
\lambda_{\sigma}v_{\sigma}^2
-\frac{\mu_0^3}{v_{\sigma}}+\mu_{3\sigma}v_{\sigma}
-\sum_{i}\left(
\frac{\mu_{i,\sigma}}{v_{\sigma}}\langle O_i\rangle
-\lambda_{i,\sigma}\langle O_i\rangle
\right)
 \,
\label{mu}
\end{eqnarray}
where $v_{\sigma}\equiv\langle S_{\sigma}\rangle$ is the VEV of
$S_\sigma$. In order to ensure the stability of the dark matter
candidate $S_D$, we require that $S_D$ does not obtain a nonzero VEV,
$\langle S_D\rangle=0$, namely $CP$ is not broken by the singlet
fields.  It follows that after the SSB, although $P$ and $CP$ are both
broken, there is a $Z_2$ type of discrete symmetry on $S_D$ remaining
in the gauge-singlet sector. The discrete symmetry is induced from the
original $CP$ symmetry.

In the limit that $v_L \simeq 0$ and $\kappa_2 \ll \kappa_1$
which comes from the phenomenology of neutrino masses and neutral meson
mixings, the mass eigenstates for the Higgs bidoublet
and triplets approximately coincide with the corresponding flavor
eigenstates. The mass terms  for the Higgs bosons and gauge bosons are
listed in  Table \ref{mass}. There is only one light SM-like
Higgs $h^0$ from the real part of $\phi^0_1$, the mass of all the other scalars are
set by $v_R$ which can be very heavy. 
\begin{table}[thb]
\begin{center}
\begin{tabular}{|l|l||l|l|}
\hline Particles  & Mass$^2$ & Particles  & Mass$^2$\\\hline
\hline
$h^0 = \phi_1^{0 r}$ &  $m_{h^0}^2 = 2 \lambda_1 \kappa^2$ & $H_1^{\pm} = \phi_1^{\pm}$ &  $m_{H_1^\pm}^2 = \frac{1}{2} \alpha_3 v_R^2 $ \\
$H_1^0 = \phi_2^{0 r}$ &  $m_{H_1^0}^2 = \frac{1}{2} \alpha_3 v_R^2$ & $H_R^{\pm \pm}=\delta_R^{\pm \pm}$ &   $m_{H_R^{\pm \pm}}^2 = 2 \rho_2 v_R^2$ \\
$A_1^0 = - \phi_2^{0 i}$ &  $m_{A_1^0}^2 = \frac{1}{2} \alpha_3 v_R^2$
& $H_L^{\pm}=\delta_L^{\pm}$ &   $m_{H_L^{\pm}}^2 = \frac{1}{2} (\rho_3 - 2 \rho_1) v_R^2$ \\
$H_2^0 = \delta_R^{0 r}$ &  $m_{H_2^0}^2 = 2 \rho_1 v_R^2 $ &
$H_L^{\pm \pm}=\delta_L^{\pm \pm}$
& $m_{H_L^{\pm \pm}}^2 = \frac{1}{2} (\rho_3 - 2 \rho_1) v_R^2$ \\
$H_3^0=\delta_L^{0r}$&   $m_{H_3^0}^2 = \frac{1}{2} (\rho_3 - 2 \rho_1)
v_R^2 $ & $A_L^0=\delta_L^{0i}$& $m_{A_L^0}^2 = \frac{1}{2} (\rho_3 - 2
\rho_1) v_R^2 $\\ \hline \hline  $Z_1$ & $m_{Z_1}^2 =
m_{W_1}^2\sec^2\theta_W$
&$W_1^\pm = W_L^\pm$ & $m_{W_1}^2 = g^2\kappa^2/4$ \\
$Z_2$ & $m_{Z_2}^2 = \frac{g^2 v_R^2 \cos^2 \theta_W}{\cos 2
\theta_W}$ &$W_2^\pm = W_R^\pm$ & $m_{W_2}^2 = g^2 v_R^2/2$
\\\hline
\end{tabular}
\end{center} \caption{The mass spectrum for the Higgs and  gauge
bosons in the LR symmetric model in the limit $v_L \simeq 0$ and $\kappa_2 \ll \kappa_1$. $\phi^{0r}_i$ and  $\phi^{0i}_i$ stand for real and imaginary component of $\phi^{0}_i$. The gauge boson $Z_1(W_1)$ corresponds to the
$Z(W)$ boson in the SM.
}
\label{mass}
\end{table}

The mass terms for $S_D$ and $S_\sigma$   are given by
\begin{eqnarray}
M^2_{D}&=&(\tilde\lambda_{\sigma}+\frac{1}{2}\tilde\lambda_{\sigma S})v_{\sigma}^2
-(\tilde\mu_{\sigma S}+\tilde\mu_{3\sigma})v_{\sigma}+\tilde\mu_{\sigma}^2
+\frac{\tilde\mu_0^3}{v_{\sigma}}+\frac{\tilde\mu_{i,\sigma}}{v_{\sigma}}\langle O_i\rangle \, , 
\nonumber\\
M^2_{\sigma}&=&2\lambda_{\sigma}v_{\sigma}^2+\left(\tilde\mu_{3\sigma}+3\tilde\mu_{\sigma S}\right)v_{\sigma}
+\frac{\tilde\mu_0^3}{v_{\sigma}}+\frac{\tilde\mu_{i,\sigma}}{v_{\sigma}}\langle O_i\rangle \, , 
\nonumber\\
M^2_{\sigma\phi_1^{0r}}&=&\kappa(-\tilde\mu_{1,\sigma}+\lambda_{1,\sigma}v_{\sigma}) \, , 
\nonumber\\
M^2_{\sigma\phi_2^{0r}}&=&2\kappa(-\tilde\mu_{2,\sigma}+\lambda_{2,\sigma}v_{\sigma})\, ,
\nonumber\\
M^2_{\sigma\delta_R^{0r}}&=&v_R(-\tilde\mu_{3,\sigma}+\lambda_{3,\sigma}v_{\sigma})\, ,
\label{md}
\end{eqnarray}
where $M^2_{\sigma\phi_1^{0r}}$, $M^2_{\sigma\phi_2^{0r}}$ and
$M^2_{\sigma\delta_R^{0r}}$ denote the mixing between singlet $S_{\sigma}$ and
the other three neutral Higgs bosons.  From the Lagrangian in
Eq. (\ref{VsComponents}) one can easily obtain the interaction terms among the
scalars. Some of the relevant cubic and quartic scalar interactions are listed in
Table \ref{interaction}.
\begin{table}[htb]
\begin{center}
\begin{tabular}{|c|c||c|c||c|c|c|c|}
\hline Interaction & Coupling& Interaction & Coupling & Interaction & Coupling & Interaction&  Coupling\\
\hline 
$S_D S_D S_{\sigma}S_{\sigma}$ 	&$2\lambda_{\sigma D}$ 
&$S_DS_Dh^0$			&$\lambda_{1,D}\kappa$
&$S_D S_D S_{\sigma}$		&$2(\mu_{\sigma D}+\lambda_{\sigma D} v_{\sigma})$
&$S_DS_DH_2^0$			&$\lambda_{3,D} v_R$
\\
$S_DS_DHH^*$			&$\lambda_{1,D}$	
&$S_{\sigma}S_{\sigma} h^0$	&$-\mu_{1,\sigma}+\lambda_{1,\sigma}\kappa$		
&$HH^*S_{\sigma}$		&$-\mu_{1,\sigma}+\lambda_{1,\sigma} v_{\sigma}$		
&$S_{\sigma}S_{\sigma} H_2^0$	&$\lambda_{3,\sigma} v_R$
\\
$S_DS_Dh^0H_1^0$		&$2\lambda_{2,D}$
&$S_DS_DH_1^0$			&$2\lambda_{2,D}\kappa$
&$h^0H_1^0S_{\sigma}$		&$2(-\mu_{2,\sigma}+\lambda_{2,\sigma}v_{\sigma})$
&$S_{\sigma}S_{\sigma}S_{\sigma}$ 	& $6\lambda_\sigma v_{\sigma}$
\\
$S_DS_D\Delta\Delta^*$		&$\lambda_{3,D}$
&$S_{\sigma}S_{\sigma} H_1^0$	&$2(-\mu_{2,\sigma}+\lambda_{2,\sigma}\kappa)$
&$\Delta\Delta^*S_{\sigma}$	&$-\mu_{3,\sigma}+\lambda_{3,\sigma} v_{\sigma}$& $h^0h^0 H^0_2$& $\alpha_1 v_R$
\\
\hline
\end{tabular}
\label{interaction}
\end{center}
\caption{The cubic and quartic scalar couplings between
Higgs singlets and multiplets, where $HH^*$ stands for any states of   
$\left\{h^0h^0,H_1^0H_1^0,A_1^0A_1^0,H_1^+H_1^-\right\}$
and
$\Delta\Delta^*$ stands for any states of $\left\{H_L^0H_L^{0},A_L^0A_L^0,H_L^+H_L^-,H_L^{++}
H_L^{--},H_2^0H_2^0,H_R^{++}H_R^{--}\right\}$. 
} \label{coupling}
\end{table}

\section{DM in the LR symmetric model}\label{DM-in-model}

There are a number of free parameters in the model such as the coefficients in
the potentials and the VEV for $S_\sigma$.  As shown in Eq. (\ref{md}), the
mass of $S_D$ is related to two energy scales $v_{\sigma}$ and $v_R$ since
$\langle O_3\rangle\sim v_R^2$. In the minimal LR model with the spontaneous
$CP$ violation, the VEV $v_R$ of the right-handed Higgs triplet $\Delta_R$ is
subjected to strong constraints from the $K$, $B$ meson mixings
\cite{Beall:1981ze,Mohapatra:1983ae,Gilman:1983ce,Chang:1984hr,Hou:1985ur,Basecq:1985cr,Ecker:1985vv,Ball:1999mb}
as well as low energy electroweak interactions
\cite{Langacker:1989xa,Barenboim:1996nd,Sahoo:2005wb}. The kaon mass
difference and the indirect $CP$ violation quantity $\epsilon_K$ set a bound
for $v_R$ around $10$ TeV
\cite{Barenboim:1996nd,Pospelov:1996fq,Zhang:2007fn}. For a successful cold DM
candidate $S_D$, its mass should be roughly between 10 GeV and a few TeV for
annihilation cross section of approximately weak strength.  For simplicity
here we consider a case in which $v_\sigma$ is heavy $v_\sigma \sim v_R\sim
10$ TeV, and before the SSB the Lagrangian has an approximate global $U(1)$
symmetry on $S$, i.e. under $S\to e^{i\delta}S$, which suppresses some of the
parameters, namely 
\begin{eqnarray}
\tilde{\mu}_0, \ \tilde{\mu}_\sigma,\ \tilde{\mu}_{\sigma S},
\ \tilde{\mu}_{3\sigma},\ \tilde{\mu}_{i,\sigma}\ll v_\sigma ,
\nonumber\\
\tilde{\lambda}_{\sigma S}, \ \tilde{\lambda}_{\sigma},\ \tilde{\lambda}_{i,\sigma}\ll 1 ,
\label{U1symmetry}
\end{eqnarray}
which leads to a relatively light $S_D$ in comparison with $v_R$, as it is the
would-be Goldstone boson in the limit of exact global $U(1)$ symmetry.  For a
light $S_D$ with $M_D\lesssim \mathcal{O}(1$TeV), without significant
fine-tuning, one needs $\tilde{\mu}_{3, \sigma} / v_\sigma \lesssim 0.01$ from
Eq. (\ref{md}). It follows from Eq. (\ref{U1symmetry}) that the coefficients
for the quartic couplings and mixing terms in the potential $\mathcal{V}_S$
are roughly at the same order of magnitude
\begin{eqnarray}
\lambda_{\sigma}\simeq\lambda_D\simeq\lambda_{\sigma D},\nonumber\\
\lambda_{i,\sigma}\simeq\lambda_{i,D}.
\end{eqnarray}
One of the implications of this parameter region is that the mixing
between $S_\sigma$ and the SM-like Higgs $h^0$ will be small. This is because
the mixing angle $\theta$ is proportional to
\begin{equation}
\tan2\theta\simeq
\frac{\kappa(-\tilde\mu_{1,\sigma}+\tilde\lambda_{1,\sigma}v_{\sigma})}{
2\lambda_\sigma v_\sigma^2 -2\lambda_1 \kappa^2
}\sim \mathcal{O}(\frac{\kappa}{v_\sigma}) .
\end{equation}
Thus the constraints from the precision electroweak data
from LEP experiments becomes weak. There are of course other
possible parameter regions. However, one will see in the next section
that the parameter space corresponding to the approximate global $U(1)$
symmetry leads to the correct magnitude of the relic dark matter
density.

\subsection{Annihilation cross section}

The relic density of the gauge-singlet DM $S_D$ can be calculated from
the annihilation cross sections which depend largely on mass spectrum
of the particles in the model, especially the mass of the DM
candidate.
$S_D$ can be very light. For $3 \,{\rm GeV} \lesssim m_D \lesssim 8$ GeV which
is consistent with the recent DAMA results \cite{Bernabei:2008yi}. $S_D$ pairs
can only annihilate to light fermion pairs through intermediate SM-like Higgs
boson $h^0$. In this case, there is an one to one correspondence between the
DM relic density and the WIMP-nucleon elastic scattering cross section. The
ratio between the two only depends on the mass of DM
\cite{Burgess:2000yq,Andreas:2008xy}. In order to satisfy both the DM relic
density $0.105 \leq\Omega_{DM} h^2 \leq 0.117$ \cite{Yao:2006px} and the
WIMP-nucleon elastic scattering cross section in the range $3\times
10^{-41}\rm{cm}^2 \alt\sigma^{SI}_{n} \alt5\times 10^{-39}\rm{cm}^2 $ reported by DAMA
\cite{Bernabei:2008yi}, a large $h^0 S_D S_D$ coupling is inevitable, which
may cause the invisible decay of $h^0$ produced at LHC \cite{Andreas:2008xy}.

\begin{figure}[t]\begin{center}
\includegraphics[scale=0.8]{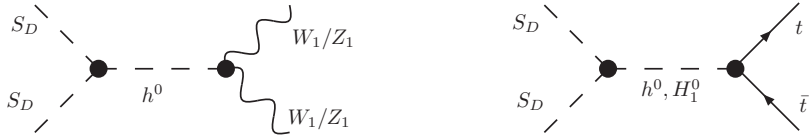}
\end{center}
\caption{Feynman diagrams for two DM candidate $S_D$ annihilating into $W_1 W_1 /Z_1 Z_1 $
and $t\bar t$ final states. } \label{WZT}\end{figure}

\begin{figure}[t]\begin{center}
\includegraphics[scale=0.8]{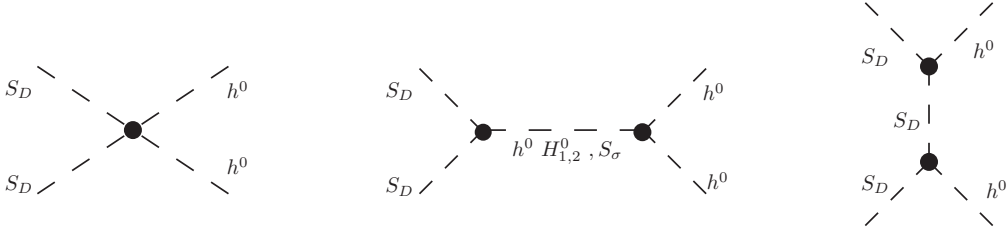}
\end{center}
\caption{Feynman diagrams for two DM candidate $S_D$ annihilating into two SM-like Higgs
bosons.} \label{h^0h^0}\end{figure}

Here we consider a different parameter rang in which $S_D$ is heavier than
the SM-like Higgs and in a mass range $200 \leq m_D \leq 500$ GeV which
can be covered by the CDMS and other experiments.
Since we assume $v_{\sigma} \sim v_R \sim 10$ TeV, most of the scalars
are heavy except for the SM-like one.  In this case, the possible
annihilation products are $h^0 h^0$, $W_1 W_1 / Z_1 Z_1$ and fermion
pairs $ q \bar q$, as shown in Fig.  \ref{WZT} and
Fig.  \ref{h^0h^0}. The $q\bar q$ final states are dominated by the heavy
$t-$ quarks since the Yukawa coupling is the largest.
For $W_1W_1/Z_1 Z_1$ finale states the only possible intermediate state is
$h^0$.  
For $q\bar q$ final states, the intermediate particles can be
$h^0$ and $H_1^0$. But $H^0_1$ contribution is negligible  as $m_{H_1^0} \gg m_{h^0}$.
Since $H^0_{2,3}$ have nonzero $B-L$ charge they can only couple to Majorana neutrinos.
For a high $v_R$ around a few TeV, the right-handed neutrinos are also heavy, which  
cannot appear in the final states.
Thus in our model the dominant annihilation processes in Fig.  \ref{WZT}
are the same as in the minimal extension of SM with a
gauge-singlet \cite{McDonald:1993ex,Burgess:2000yq}.
For  $h^0h^0$ final states, the $s-$channel involves $h^0$, $H^0_{1,2}$
and $S_\sigma$ while the $t-$channel involves $h^0$ only.

The relevant annihilation cross sections for  Fig.  \ref{WZT} are given by
\begin{eqnarray}
(4 E_1E_2\sigma v)_{W_1W_1}
&=&\frac{\lambda_{1,D}^2}{8\pi}(1-\frac{m_{h^0}^2}{s})^{-2}
\left(1-4\frac{m_{W_1}^2}{s}+12\frac{m_{W_1}^4}{s^2}\right)(1-\frac{4m_{W_1}^2}{s})^{\frac{1}{2}}\label{WW},
\nonumber\\
(4 E_1E_2\sigma v)_{Z_1 Z_1}
&=&\frac{\lambda_{1,D}^2}{16\pi}(1-\frac{m_{h^0}^2}{s})^{-2}\left(1-4\frac{m_{Z_1}^2}{s}
+12\frac{m_{Z_1}^4}{s^2}\right)(1-\frac{4m_{Z_1}^2}{s})^{\frac{1}{2}}\label{ZZ},
\nonumber\\
(4E_1E_2\sigma v)_{tt}
&=&\frac{3\lambda_{1,D}^2}{4\pi}\frac{m_t^2}{s}(1-\frac{4m_t^2}{s})^{\frac{3}{2}}(1-\frac{m_{h^0}^2}{s})^{-2}\label{tt},
\end{eqnarray}
and that for Fig.  \ref{h^0h^0} is
\begin{eqnarray}
(4E_1E_2\sigma v)_{h^0h^0}
&=&\frac{\lambda_{1,D}^2}{16\pi}\left(1-\frac{4m_{h^0}^2}{s}\right)^{\frac{1}{2}}
\times\left[\left(\frac{s-4m_{h^0}^2}{s-m_{h^0}^2}-\frac{m^2_{\sigma}}{s-m^2_{\sigma}}
-\frac{\alpha_1\lambda_{3,D}}{\lambda_{1,D}}\frac{v_R^2}{s-m^2_{H_2^0}}\right)^2\right.\nonumber\\
&&+4\lambda_{1,D}\frac{\kappa^2}{s-2m_{h^0}^2-2m_D^2}\left.
\left(\frac{s-4m_{h^0}^2}{s-m_{h^0}^2}-\frac{m^2_{\sigma}}{s-m^2_{\sigma}}
-\frac{\alpha_1\lambda_{3,D}}{\lambda_{1,D}}\frac{v_R^2}{s-m^2_{H_2^0}}\right)Y(\xi)\right.\nonumber\\
&&+\left.2\lambda_{1,D}^2\left(\frac{\kappa^4}{(m_D^2-m_{h^0}^2)^2}+\frac{4\kappa^4}{(s-2m_{h^0}^2-2m_D^2)^2}Y(\xi)\right)\right]
\label{hh}
\end{eqnarray}
where $s$ is the squared center-of-mass energy. $E_1$ and $E_2$ are the energies of 
the incidental particles. The quantity $Y$ is defined as $Y(\xi)\equiv\mbox{arctanh}(\xi)/\xi$ with
$\xi \equiv\sqrt{(s-4m_{h^0}^2)(s-4m_D^2)}/(s-2m_{h^0}^2-2m_D^2)$.
For the cross section $(4E_1E_2\sigma v)_{h^0h^0}$ in Eq. (\ref{hh}),
the $H_2^0 h^0 h^0$ scalar coupling  $\alpha_1 v_R$  has been used
 \cite{Guo:2008hy}. 

\subsection{Constraints from The DM Relic density}

The thermal-average of the annihilation cross section times the
relative velocity $\langle \sigma v \rangle$ is a key quantity in the
determination of the cosmic relic abundance of $S_D$. For
nonrelativistic gases, $\langle \sigma v \rangle$ can be expanded in
powers of relative velocity and  $x^{-1}$ ($x \equiv m_D/T$). To the first order 
$\langle \sigma v \rangle \simeq 
\sigma_0 x^{-n}$, where $n=0(1)$ for $s(p)$-wave annihilation process \cite{Kolb:1990vq}. 
The general formula for $\langle\sigma v\rangle$ is given by  \cite{Srednicki:1988ce}
\begin{eqnarray}
\langle \sigma v \rangle = \sigma_0 x^{-n} = \frac{1}{m_D^2} \left [
\omega - \frac{3}{2} (2 \omega - \omega ')x^{-1} + \ldots
\right]_{s/4m_D^2=1} \; , \label{Expand}
\end{eqnarray}
where $\omega \equiv E_1 E_2 \sigma v$, and the prime denotes derivative with
respect to $s/4m_D^2$. 
$\omega$ and its derivative are all to be evaluated at $s/4m_D^2=1$.
The final DM density $\Omega_{DM} h^2$ is given by
 \cite{Kolb:1990vq}
\begin{eqnarray}
\Omega_{DM} h^2 = 1.07 \times 10^9 \, \frac{(n+1)
x_f^{n+1}}{g_*^{1/2} M_{Pl} \, \sigma_0} \, {\rm GeV}^{-1}
\label{nin}
\end{eqnarray}
with
\begin{eqnarray}
x_f & = & {\rm ln}[0.038(n+1)(g_{D}/g_*^{1/2}) M_{Pl}  \, m_D \,
\sigma_0] 
\nonumber \\ 
&&- (n+ 1/2)\, {\rm ln} \{ {\rm ln}
[0.038(n+1)(g_{D}/g_*^{1/2}) M_{Pl} \, m_D \, \sigma_0] \} \; ,
\label{xf}
\end{eqnarray}
where $M_{Pl} = 1.22 \times 10^{19}$ GeV and $g_D = 1$ is the
internal degree of freedom of $S_D$. $g_*$ is the total number of
effectively relativistic degrees of freedom at the time of
freeze-out. For particles  playing the role of cold
DM, the relevant freeze-out temperature is $x_f = m_D/T_f \sim 25$.
Since we consider the range $ 200 \,{\rm GeV} \leq m_D \leq 500\, {\rm GeV}$ in  
our analysis, we obtain $g_* = 345/4$.

\begin{figure}[htb]\begin{center}
\includegraphics[scale=0.25]{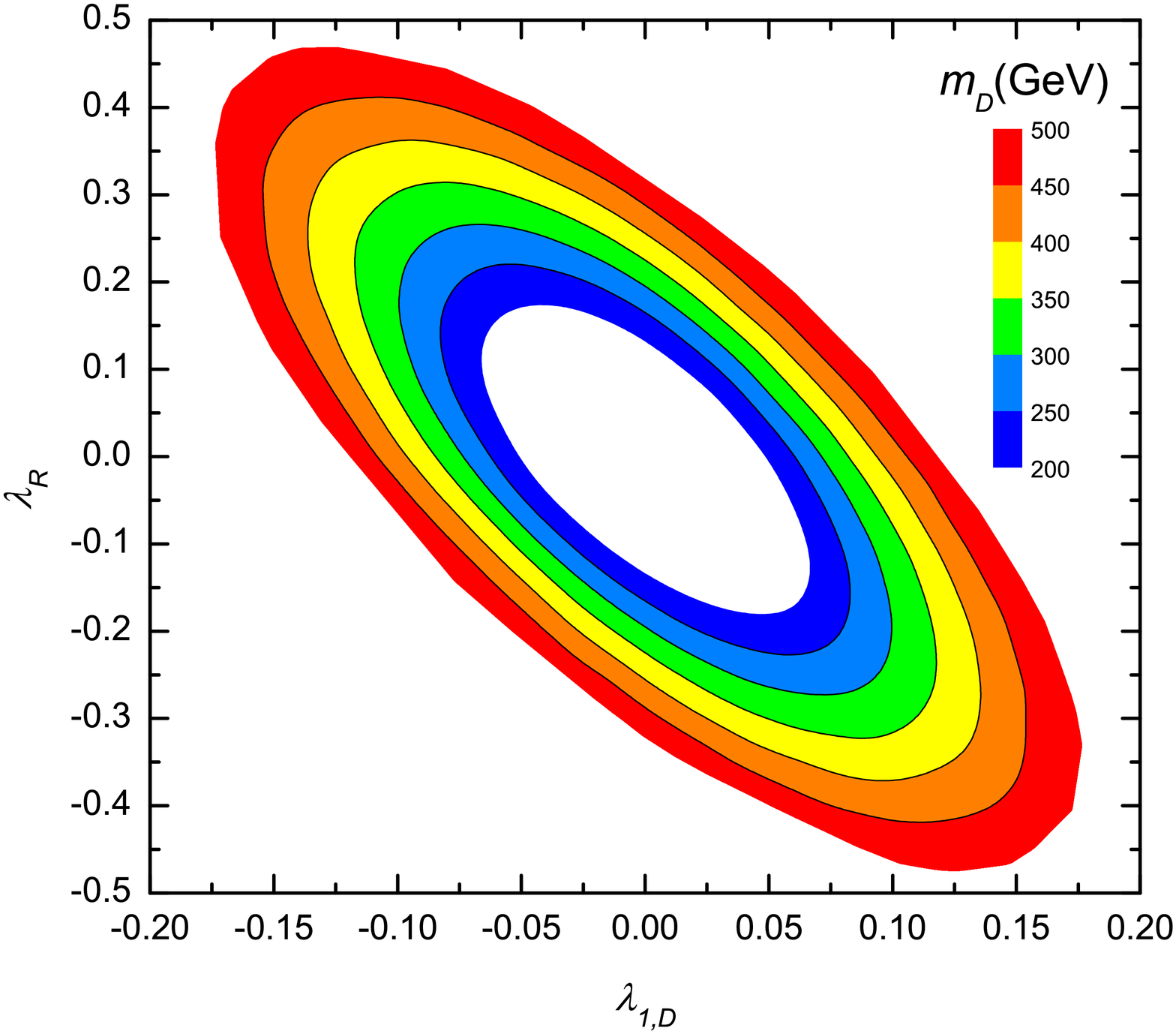}
\includegraphics[scale=0.25]{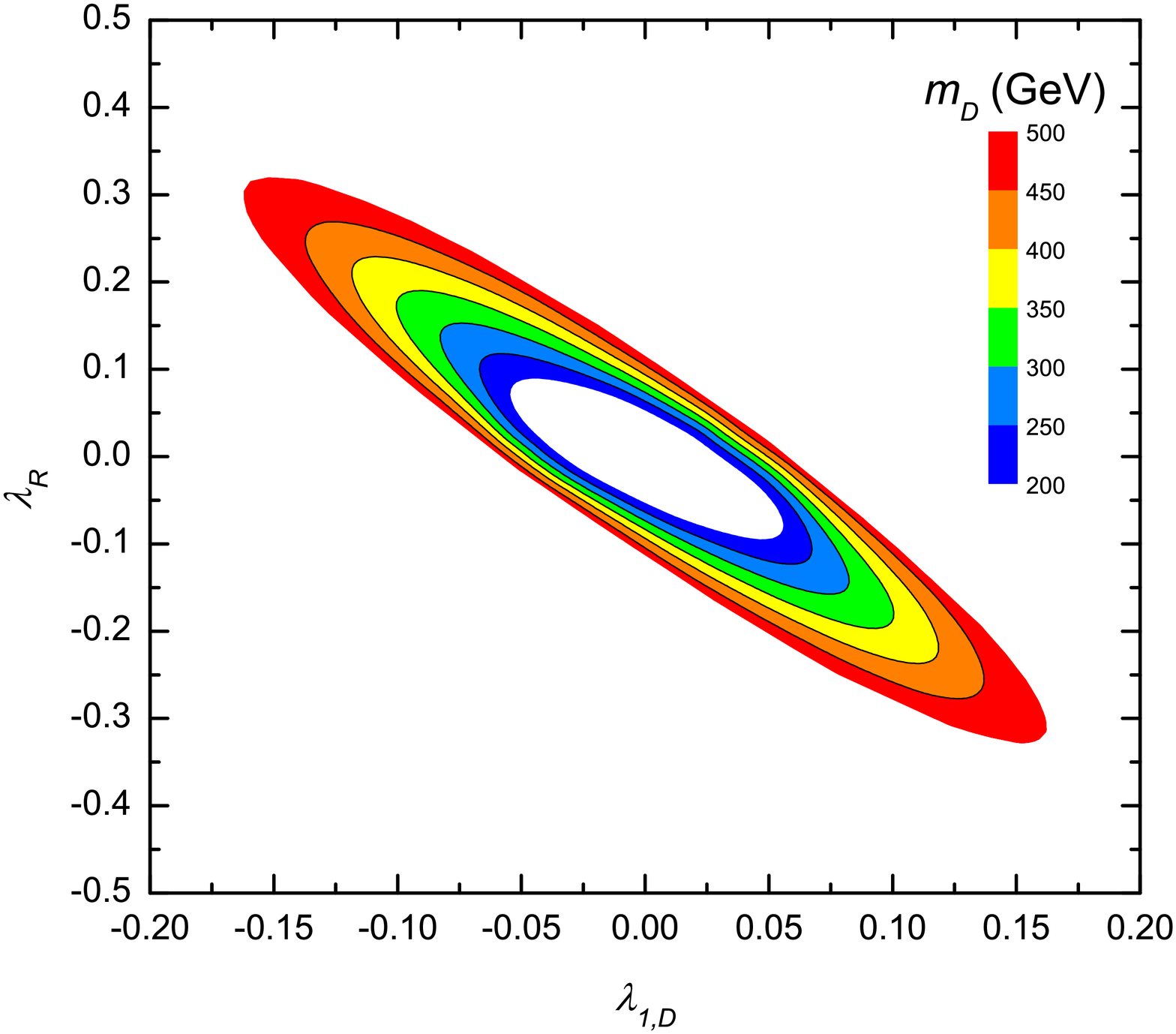}
\end{center} %
\caption{ The allowed region of $\lambda_{1,D}$ and $\lambda_R$ for
different $m_D$  from  DM relic density.
The left panel corresponds to the annihilation involving only one Higss bidoublet.
The right panel corresponds to the annihilation involving two Higgs bidoublet. See the text
for detailed explanation.  
} \label{space}
\end{figure}

The total annihilation cross section $\omega$ is
\begin{eqnarray} \omega = (E_1E_2\sigma v)_{h^0 h^0} +
(E_1E_2\sigma v)_{W_1 W_1}+ (E_1E_2\sigma v)_{Z_1 Z_1}+(E_1E_2\sigma
v)_{tt} \;. \label{omega}
\end{eqnarray}
From Eq. (\ref{hh}) there are seven unknown parameters enter the
expression of total annihilation cross section, namely, $m_{h^0}$,
$m_D$,   $\lambda_{1,D}$, $\alpha_1\lambda_{3,D}$, $m_\sigma^2$, 
$m_{H_2^0}^2$ and $v_R$. But $\omega$ is highly 
insensitive to $m_\sigma$ and $m_{H_2^0}^2$ when $s_\sigma$ and $H^0_{2}$ masses
are around $v_R$. Furthermore the mass of $H^0_{2}$ can be related to 
$v_R$ through $m_{H_2^0}^2 \approx 2 \rho_1 v_R^2$ as it is shown in Table \ref{mass}.  
Thus only four
parameters 
$$m_{h^0}, \ m_D, \ \lambda_{1,D} \mbox{ and } \lambda_R \equiv
\alpha_1\lambda_{3,D}/(2\rho_1)$$ are relevant to our numerical analysis. In
numerical calculations, we fix the mass of the SM-like Higgs to $m_{h^0} = 120$
GeV, and perform a numerical scan over the parameters $\lambda_R$ and
$\lambda_{1,D}$ for the mass range $200 \,{\rm GeV} \leq m_D \leq 500\, {\rm
  GeV}$.  The allowed parameter space is shown in Fig.~\ref{space} (left
panel), which gives an allowed range
$$-0.18 \alt \lambda_{1, D} \alt 0.18 \ \mbox{ and } \ -0.48 \alt \lambda_R \alt 0.47.$$ 
The central region of this figure is
excluded since these points cannot provide large enough annihilation cross
section to give the desired relic abundance. For such a mass range of $S_D$,
without significant fine-tuning, one needs $\tilde{\lambda} \lesssim 0.01$ and
$\tilde{\mu}/v_\sigma \lesssim 0.01$ from Eq. (\ref{md}), where
$\tilde{\lambda}$ and $\tilde{\mu}$ denote the corresponding parameters in
Eq. (\ref{U1symmetry}). Since the approximate global symmetry $U(1)$ requires
$\tilde{\lambda} \ll \lambda_{1,D}$, the region near $\lambda_{1, D} =0$ in
Fig.  \ref{space} is disfavored.

\subsection{Predictions for the DM direct detection experiments}

The current DM direct detection experiments, such as the
CDMS \cite{Ahmed:2008eu} and XENON \cite{Angle:2007uj}, have imposed  strong
constraints on the WIMP-nuclen elastic scattering cross section for a wide
range of DM mass. In
our model, the DM candidate $S_D$ interacts with nucleus
${\cal N}$ through  Yukawa couplings interactions. For scalar
interactions, the spin-independent elastic scattering cross section
on a nucleus ${\cal N}$ is given by  \cite{Bertone:2004pz,Jungman:1995df}
\begin{eqnarray}
\sigma_{\cal N} = \frac{4 M^2({\cal N})}{\pi} (Z f_p + (A-Z) f_n)^2
\;,
\end{eqnarray}
where $M({\cal N}) = m_D M_{\cal N}/(m_D + M_{\cal N})$ and $M_{\cal
N}$ is the target nucleus mass. $Z$ and $A-Z$ are the numbers of
protons and neutrons in the nucleus. $f_{p,n}$ is the
coupling between WIMP and  protons or neutrons, given by
\begin{eqnarray}
f_{p,n}= \sum_{q=u,d,s} f_{Tq}^{(p,n)} a_q \frac{m_{p,n}}{m_q} +
\frac{2}{27} f_{TG}^{(p,n)}  \sum_{q=c,b,t} a_q \frac{m_{p,n}}{m_q},
\label{fn}
\end{eqnarray}
where $f_{Tu}^{(p)}=0.020 \pm 0.004$, $f_{Td}^{(p)}=0.026 \pm 0.005$,
$f_{Ts}^{(p)}=0.118 \pm 0.062$, $f_{Tu}^{(n)}=0.014 \pm 0.003$,
$f_{Td}^{(n)}=0.036 \pm 0.008$ and $f_{Ts}^{(n)}=0.118 \pm 0.062$
 \cite{Ellis:2000ds}. The coupling $f_{TG}^{(p,n)}$ between WIMP and gluons from
heavy quark loops is obtained from
\begin{eqnarray}
f_{TG}^{(p,n)} = 1 - \sum_{q=u,d,s} f_{Tq}^{(p,n)} \;.
\end{eqnarray}
Traditionally, the results of WIMP-nucleus elastic
scattering experiments are presented in the form of a normalized
WIMP-nucleon scattering cross section $\sigma_{n}^{SI}$ in the
spin-independent case, which is straightforward
\begin{eqnarray}
\sigma_{n}^{SI} =\frac{1}{A^2} \frac{M^2(n)}{ M^2({\cal N})}
\sigma_{\cal N} \; ,
\end{eqnarray}
where $M(n) = m_D M_{n}/(m_D + M_{n})$ is the reduced mass of the
nucleon, and $M_n = m_{p, n}$ denotes the nucleon mass. Because of
$f_p \approx f_n$ in our model
\begin{eqnarray}
\sigma_n^{SI} \approx \frac{4 M^2_n}{\pi} f_{n}^2.
\end{eqnarray}
The present bounds on the WIMP-nucleon elastic scattering cross
section are  $\sigma_{n}^{exp} \leq 8 \times 10^{-44}\,
{\rm cm^2}\sim 2 \times 10^{-43}\,
{\rm cm^2}\ $ from the CDMS  \cite{Ahmed:2008eu} and
$\sigma_{n}^{exp} \leq 2 \times 10^{-44} \, {\rm cm^2} \sim 4.3 \times 10^{-43} \, {\rm cm^2} $ for 
from the XENON  \cite{Angle:2007uj} for the DM  mass range $200\sim 500$ GeV.

\begin{figure}[h]\begin{center}
\includegraphics[scale=0.25]{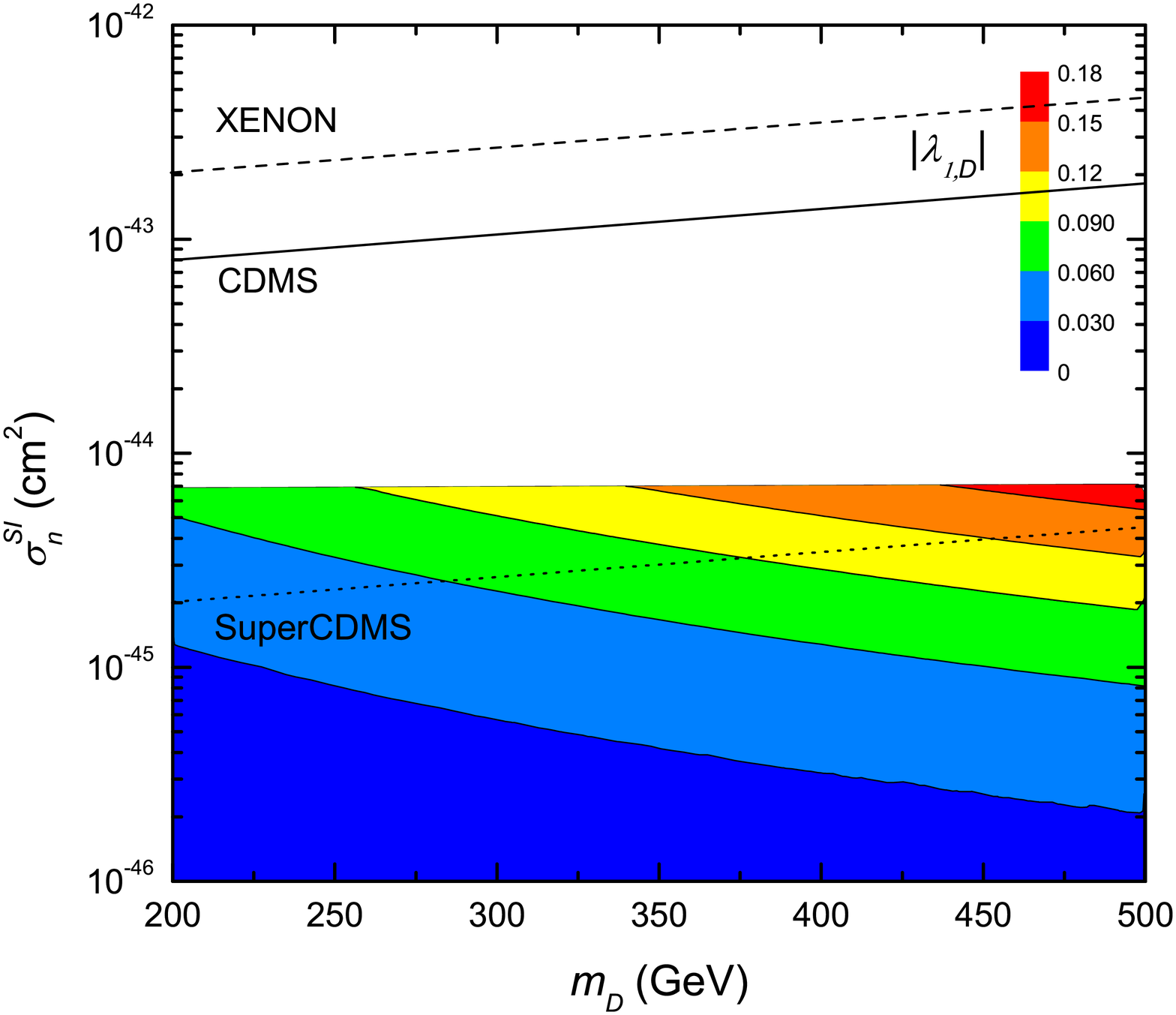}
\includegraphics[scale=0.25]{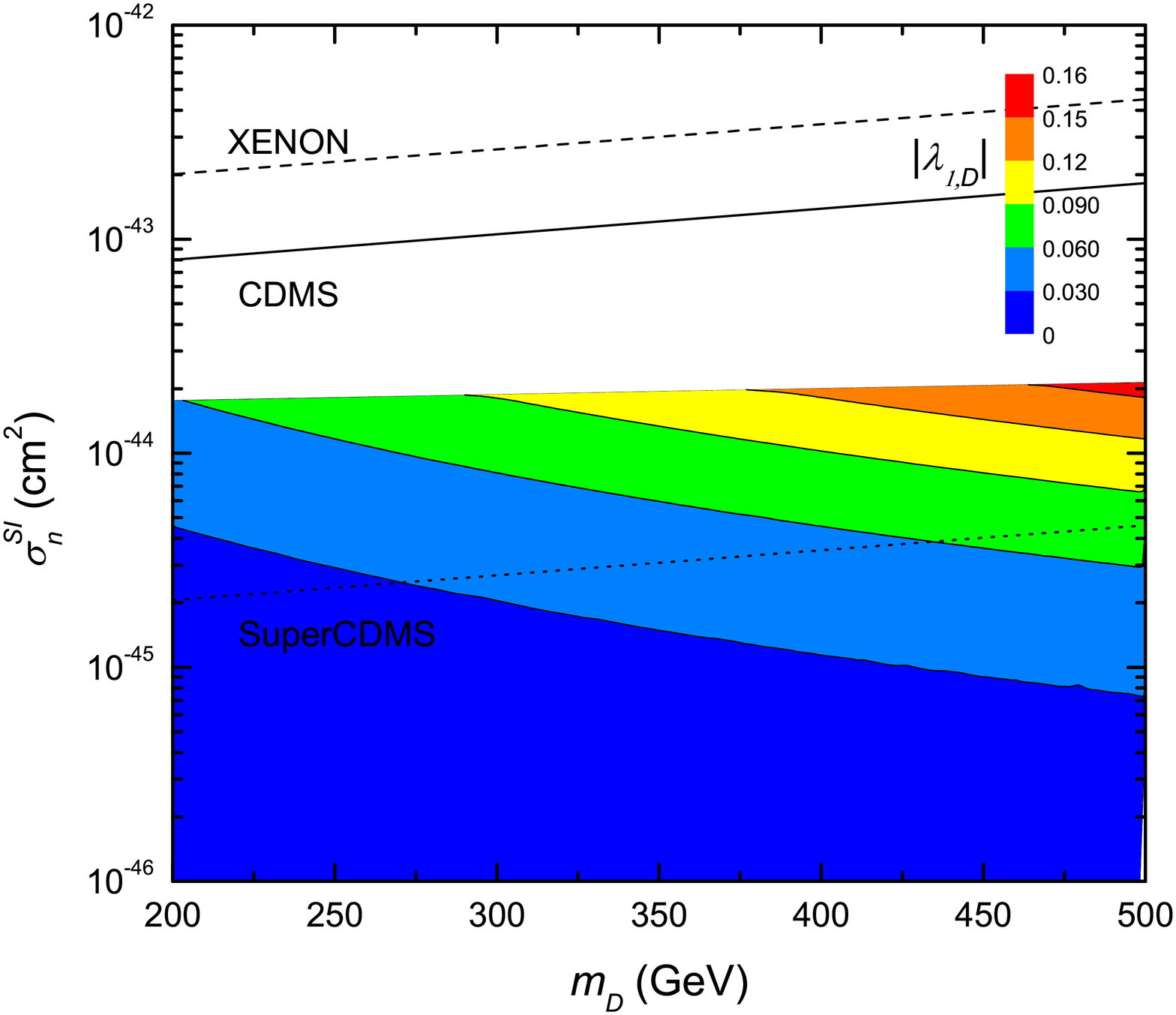}
\end{center}
\caption{Predicted region of the spin-independent WIMP-nucleon
elastic scattering cross section $\sigma_{n}^{SI}$ as functions of
$m_D$ and $\lambda_{1, D}$.  The dashed line and solid line
denote the present experimental upper bounds from the XENON and
CDMS, respectively. The dotted line indicates the sensitivity of the
future SuperCDMS  \cite{Akerib:2006rr}. 
The left panel corresponds to the annihilation involving only one Higss bidoublet.
The right panel corresponds to the annihilation involving two Higgs bidoublet with 
the assumption of $y^{\eta_1}_{q}\simeq y^{\eta_2}_{q} \simeq y^h_q$. 
See the text
for detailed explanation.  
}\label{direct}
\end{figure}

The DM candidate $S_D$ interacts with nucleus
${\cal N}$ through their couplings with quarks by exchanging Higgs
bosons $h^0$ and $H_1^0$. Because  ${H_1^0}$ is much heavier than $h^0$, the
main contribution comes from the $h^0$-exchange only. In this
case, the WIMP-quark coupling $a_q$ in Eq. (\ref{fn}) is given by
\begin{eqnarray}
a_q =  \frac{\lambda_{1,D} \, y^h_q \, \kappa }{2 \sqrt{2} \,
m_D \, m_{h^0}^2} \, \label{aq}
\end{eqnarray}
where $y^h_q$ ($q= u, d, s, c, b, t$) denotes the Yukawa
coupling of the SM-like Higgs to the quarks with
$ y^h_q \kappa / \sqrt{2}= m_q$. Using the allowed ranges for $\lambda_{1,
D}$  we make predictions for  the spin-independent WIMP-nucleon elastic
scattering cross section $\sigma_{n}^{SI}$. The numerical results
are shown in Fig.  \ref{direct} (left panel). One  finds
$\sigma_{n}^{SI} \lesssim 7\times 10^{-45} {\rm cm}^2$ for $200
\,{\rm GeV} \leq m_D \leq 500\, {\rm GeV}$,  which is far bellow 
the current experimental upper bounds. Nevertheless the future SuperCDMS (Phase
A) experiment  \cite{Akerib:2006rr} is able to  cover  part of the allowed
parameter space, especially in the small $m_D$ region.

\section{Contributions from the other Higgs-bidoublet}\label{DM-2HBLR}

In this section, we generalize the previous discussions to the case in
which the other bidoublet $\chi$ mixes significantly with $\phi$ and
$\Delta_{L,R}$.  In this case the SCPV can be easily
realized \cite{Wu:2007kt,Wu:2007gb,wang2009}. Comparing with the previous case, the main
differences are that there could be more scalar particles entering the
DM annihilation and scattering processes. Furthermore, the new contributions from
these particles may modify the correlation between the DM annihilation
and WIMP-nucleus elastic scattering cross sections, which leads to significantly
different predictions from the  other gauge-singlet scalar DM models and 
the previous discussions.

As shown in Eq. (1), the second Higgs bidoublet $\chi$ contains two
neutral Higgs particles $\chi_{1,2}^0$. After the SSB,
$\chi_{1,2}^0$ may obtain  VEVs $w_{1,2}/\sqrt{2}$. The squared 
sum of all the VEVs including $\kappa_{1,2}$ should still lead to 
$v=(|\kappa_1|^2 + |\kappa_1|^2 +|w_1|^2+ |w_2|^2)^{1/2}  \approx 246$GeV. 
In the physical basis, some of the Higgs bosons from $\chi$ could be
light around electroweak scale. The number of the light Higgs depends on
the Higgs potential. In most cases, there are two more light neutral
Higgs $\eta^0_{1,2}$ and one pair of light charged Higgs $\eta^\pm$ \cite{wang2009}. This
feature can be easily understand in the limit $\kappa_2 \sim w_2
\sim 0$.  In this case, one can determine $h^0$, $\eta^0_1$ and
$\eta^0_2$ from $\phi_1^0$ and $\chi_1^0$, and $\eta_{\pm}$ from the
mixing of $\phi_1^{\pm}$ and $\chi_1^{\pm}$.
The number of kinematically allowed DM annihilation processes 
depends on the masses of the relevant particles.  Here we consider a 
case in which  $S_D$ is heavier than all the light scalars
and the SM-like $h^0$ remains the lightest scalar, i.e.
$m_{h^0} \leq m_{\eta^0_{1,2}},m_{\eta^+} \leq m_D$.
The quartic interaction and the $s$-channel annihilation in Fig.  \ref{h^0h^0} now
have seven possible  final states. They are combinations of any two of the
three neutral states $(h^0,\eta^0_1,\eta^0_2)$ and charged final states
$\eta^+\eta^-$.  Note that each $s$-channel diagram in
Fig.  \ref{h^0h^0} may have $h^0$, $\eta^0_1$ and $\eta^0_2$ as
intermediate states besides $S_\sigma$ and $H^0_2$.
The $t$-channel diagram has six possible final states, due to the absence of
the cubic  scalar vertexes $S_D S_D\eta^\pm$.

The cubic coupling $S_D S_D \eta^0_{1,2}$ although can be
different from that for $S_D S_D h^0$, may not modify the correlation
between the DM annihilation and WIMP-nucleon elastic scattering cross
sections in a significant way. As it is pointed out in Ref. \cite{Andreas:2008xy}, the
ratio $R\equiv \langle \sigma v \rangle/\sigma^{SI}_{n}$ is highly insensitive to
these couplings because they cancel out largely. $R$ is only sensitive
to the mass of DM candidate and the Yukawa couplings. In the minimal scalar DM
model the Yukawa couplings are the same as that in the SM. it is shown
that the value $R$ scales as $m_D^2$ \cite{Hambye:2008bq}. For small $m_D$ around a few GeV
the value of $R$ is in agreement with the DAMA results. A large $m_D$
around a few hundred GeV corresponds to a large $R$, which indicates
that the WIMP-nucleon elastic scattering cross section may be far
bellow the current direct detection bounds.

In 2HBLR the Yukawa couplings can be significantly different from those
in the SM and the minimal LR model. Similar to the general
two-Higgs-doublet model \cite{Hall:1993ca,cheng:1987rs,wu:1994ja,Wu:1994ib,Wolfenstein:1994jw,Wu:1998nga,Wu:1999fe,Wu:2001vq} the Yukawa couplings are not simply determined
by the quark masses. This is because with the introduction of the
additional bidoublet, the fermion mass matrices and Yukawa matrices
are not proportional to each other. In general the Yukawa couplings can be
parameterized as $y_q=\sqrt{2}\xi_q m_q/v$, the factor $\xi_q$ depends
on fermion flavor $q$ and can  be different from unity. For the DM
annihilation processes, the heavy quark contribution dominates while for 
WIMP-nucleon scattering processes the light quarks are more important as
the quark mass dependence are reduced in Eq. (\ref{fn}). For a large $\xi_q$
for light quark sector it is possible that the prediction
for WIMP-nucleon elastic scattering cross section can be enhanced and
the cross section for DM annihilation still coincides with the
observed DM relic density.
%

For a concrete  numerical illustration,  we choose all the 
masses $m_{\eta^0_1}$,
$m_{\eta^0_2}$, $m_{\eta^\pm}$ at 180 GeV and keep  $m_{h^0}=120$ GeV.
%
For cubic and quartic scalar couplings, we assume they are the same as
that for the SM-like Higgs. Namely, the couplings of $S_D S_D
\eta^0_{1,2}$ and $S_D S_D \eta^0_{1,2}\eta^0_{1,2}$ are set equal to
$\lambda_{1, D} v$ and $\lambda_{1, D}$, respectively.
Similarly, the cubic scalar couplings among the light Higgs particles $h^0$,
$\eta^{0}_{1,2}$ and $\eta^{\pm}$ are set equal to $3 m_{h^0}^2/v$.  For a
comparison we consider two sets of Yukawa couplings:

\noindent{\bf i)}
All the couplings $y^h_q$, $y^{\eta_1}_{q}$, $y^{\eta_2}_{q}$
for $h^0 q\bar q$, $\eta^0_1 q\bar q$, $\eta^0_2 q\bar q$ respectively 
are  nearly the same: 
\begin{equation}
y^{\eta_1}_{q}\simeq y^{\eta_2}_{q} \simeq y^h_q .
\end{equation}
with $y^h_q=\sqrt{2}m_q/v$. In this case annihilation cross section
can be obtained simply by counting the number of new channels. In
Fig.  \ref{space} (right panel), we give the constraints on
$\lambda_{1,D}$, $\lambda_R$ for different $m_D$. It is clear that
there is a stronger constraint on the allowed parameter space, due to
the increased number of intermediate and final states.  For the
WIMP-nucleon elastic scattering process, the WIMP-quark coupling $a_q$
in Eq. (\ref{aq}) is given by
\begin{eqnarray}
a_q =  \frac{\lambda_{1,D} \, v }{2 \sqrt{2} \, m_D } \left (
\frac{y^h_q}{m^2_{h^0}} + \frac{y^{\eta_1}_q}{m^2_{\eta^0_1}}+
\frac{y^{\eta_2}_q}{m^2_{\eta^0_2}} \right ) \,. \label{aq2}
\end{eqnarray}
Using the allowed $\lambda_{1, D}$ and $m_D$ from Fig.~\ref{space}, we
calculate the spin-independent WIMP-nucleon elastic scattering cross section
$\sigma_{n}^{SI}$.  The numerical results are shown in Fig.~\ref{direct}
(right panel). We find that the $\sigma_n^{SI}$ is enlarged roughly by a
factor of three, which is however still  below the current experiment upper
bounds.

\begin{figure}[htb]\begin{center}
\includegraphics[scale=0.25]{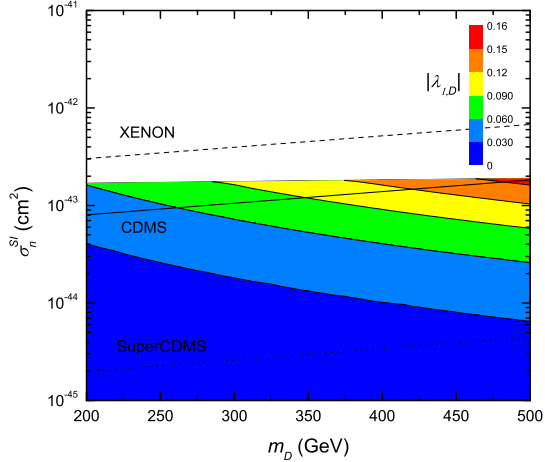}
\end{center}
\caption{Predicted region of the spin-independent WIMP-nucleon
elastic scattering cross section $\sigma_{n}^{SI}$ as functions of
$m_D$ and $\lambda_{1, D}$ in two Higgs bidoublet model with
the assumption of  $y^{\eta_1}_q \simeq y^{\eta_2}_q \simeq 10 y^h_q$ for three  light quarks. }\label{di3}
\end{figure}

\noindent{\bf ii)} The Yukawa couplings for $y^{\eta_1}_q$
and $y^{\eta_2}_q$ are significantly larger in the light quark sector
(for $q=u,d,s$)
\begin{equation}
y^{\eta_1}_q \simeq y^{\eta_2}_q \simeq 10 y^h_q .
\end{equation}
Since the annihilation process $S_D S_D \to q\bar q$ is dominated by heavy
$t$-quarks,  the enhanced Yukawa couplings $y^{\eta_1}_q$
and $y^{\eta_2}_q$ do not affect the total
annihilation cross section. Thus the DM relic density remain unchanged.  
However, the predicted WIMP-nucleon scattering cross
section $\sigma_{n}^{SI}$ will be enhanced. 
The corresponding results have been shown in Fig.~\ref{di3}. 
We find that in this case $\sigma^{SI}_n$ is enhanced by an order of
magnitude compared with the one-Higgs bidoublet case.  
The future DM direct detection experiment
SuperCDMS can cover most of the allowed parameter space.

\section{Conclusions}\label{conclusions}

In summary, we have discussed the possibility that the stability of DM can
protected by the fundamental symmetries $P$ and $CP$ of quantum field
theory. It   can be realized in the framework of a generalized LR symmetric
model which allows SCPV. The DM candidate in our model is a gauge-singlet 
which transforms under $CP$ as an ordinary complex scalar.  In this model no
extra discrete symmetry is required.
We have scanned the parameter space allowed by the relic DM density 
and  made predictions for direct detection experiments. We found that
the model was in agreement with the current measurement in a large
parameter space. 
Based on the constrained parameter space, we have made predictions for the
WIMP-nucleon spin-independent cross sections, and further studied the
correlations with the DM annihilation.  We have found that in this model the
correlation could be significantly different from other gauge-singlet DM models.
The DM-nucleon elastic scattering cross section could reach the current
experimental upper bound for large Yukawa couplings for light quarks, which
could be tested by the future experiments.

\acknowledgments{This work was supported by the National Nature
Science Foundation of China (NSFC) under the Grant No. 10475105 and
No. 10491306. W. L. Guo is supported by the China Postdoctoral
Science Foundation and the K. C. Wong Education Foundation (Hong
Kong).}

\bibliographystyle{/home/zhou/reflist/bibstyle/h-physrev4} 

\bibliography{/home/zhou/reflist/2HDM,/home/zhou/reflist/darkmatter,/home/zhou/reflist/LRmodel}

\end{document}